\documentclass[prl,twocolumn,superscriptaddress,showpacs]{revtex4}
\usepackage{epsfig}

\usepackage{graphicx} 
\usepackage{dcolumn}  
\usepackage{bm}       
\usepackage{amsfonts, amssymb}

\newcommand{\tc}{,~}

 \def\bea{\begin{eqnarray}}
 \def\eea{\end{eqnarray}}

\newcommand{\gevsq}{GeV$^2$}

\newcommand{\invs}{$s$}
\newcommand{\invt}{$t$}

\newcommand{\gep}{$G_E^p$}

\newcommand{\KLL}{$K_{_{LL}}$}
\newcommand{\KLS}{$K_{_{LS}}$}

\newcommand{\PR}{{\em Phys. Rev. }}

\newcommand{\PRL}{{\em Phys. Rev. Lett. }}

\newcommand{\NP}{{\em Nucl. Phys. }}
\newcommand{\NIM}{{\em Nucl. Instr. Meth. }}
\newcommand{\etal}{{\em et al.}}
\begin{document}

\preprint{APS/123-QED}


\title{Polarization Transfer in Proton Compton Scattering at High Momentum Transfer}

\author{D.~J.~Hamilton}
\affiliation{\mbox{University of Glasgow, Glasgow G12 8QQ, Scotland, U.K.}}
\author{V.~H.~Mamyan}
\affiliation{\mbox{Yerevan Physics Institute, Yerevan 375036,
Armenia}} \affiliation{\mbox{Thomas Jefferson National Accelerator
Facility\tc Newport News, VA 23606}}
\author{K.~A.~Aniol}
\affiliation{\mbox{California State University Los Angeles\tc Los Angeles\tc CA 90032}}
\author{J.~R.~M.~Annand}
\affiliation{\mbox{University of Glasgow, Glasgow G12 8QQ, Scotland, U.K.}}
\author{P.~Y.~Bertin}
\affiliation{\mbox{Universit\'{e} Blaise Pascal/IN2P3\tc F-63177 Aubi\`{e}re, France}}
\author{L.~Bimbot}
\affiliation{\mbox{IPN Orsay B.P. n$^\circ$1 F-91406, Orsay, France}}
\author{P.~Bosted}
\affiliation{\mbox{University of Massachusetts, Amherst, MA 01003}}
\author{J.~R.~Calarco}
\affiliation{\mbox{University of New Hampshire, Durham, NH 03824}}
\author{ A.~Camsonne}
\affiliation{\mbox{Universit\'{e} Blaise Pascal/IN2P3\tc F-63177 Aubi\`{e}re, France}}
\author{G.~C.~Chang}
\affiliation{\mbox{University of Maryland, College Park, MD 20742}}
\author{T.-H.~Chang}
\affiliation{\mbox{University of Illinois, Urbana-Champaign, IL 61801}}
\author{J.-P.~Chen}
\affiliation{\mbox{Thomas Jefferson National Accelerator Facility\tc Newport News, VA 23606}}
\author{Seonho~Choi}
\affiliation{\mbox{Temple University, Philadelphia, PA 19122}}
\author{E.~Chudakov}
\affiliation{\mbox{Thomas Jefferson National Accelerator Facility\tc Newport News, VA 23606}}
\author{A.~Danagoulian}
\affiliation{\mbox{University of Illinois, Urbana-Champaign, IL 61801}}
\author{P.~Degtyarenko}
\affiliation{\mbox{Thomas Jefferson National Accelerator Facility\tc Newport News, VA 23606}}
\author{C.~W.~de~Jager}
\affiliation{\mbox{Thomas Jefferson National Accelerator Facility\tc Newport News, VA 23606}}
\author{A.~Deur}
\affiliation{\mbox{University of Virginia, Charlottesville, VA 22901}}
\author{D.~Dutta}
\affiliation{\mbox{Duke University and TUNL, Durham, NC 27708}}
\author{K.~Egiyan}
\affiliation{\mbox{Yerevan Physics Institute, Yerevan 375036, Armenia}}
\author{H.~Gao}
\affiliation{\mbox{Duke University and TUNL, Durham, NC 27708}}
\author{F.~Garibaldi}
\affiliation{\mbox{INFN, Sezione di Sanit\'{a}
and Institute Superiore di Sanit\'{a} \tc I-00161 Rome\tc Italy}}
\author{O.~Gayou}
\affiliation{\mbox{College of William and Mary, Williamsburg, VA 23187}}
\author{R.~Gilman}
\affiliation{\mbox{Thomas Jefferson National Accelerator Facility\tc Newport News, VA 23606}}
\affiliation{\mbox{Rutgers, The State University of New Jersey\tc Piscataway, NJ 08854}}
\author{A.~Glamazdin}
\affiliation{\mbox{Kharkov Insitute of Physics and Technology\tc Kharkov 61108, Ukraine}}
\author{C.~Glashausser}
\affiliation{\mbox{Rutgers, The State University of New Jersey\tc Piscataway, NJ 08854}}
\author{J.~Gomez}
\affiliation{\mbox{Thomas Jefferson National Accelerator Facility\tc Newport News, VA 23606}}
\author{J.-O.~Hansen}
\affiliation{\mbox{Thomas Jefferson National Accelerator Facility\tc Newport News, VA 23606}}
\author{D.~Hayes}
\affiliation{\mbox{Old Dominion University, Norfolk, VA 23529}}
\author{D.~Higinbotham}
\affiliation{\mbox{Thomas Jefferson National Accelerator Facility\tc Newport News, VA 23606}}
\author{W.~Hinton}
\affiliation{\mbox{Old Dominion University, Norfolk, VA 23529}}
\author{T.~Horn}
\affiliation{\mbox{University of Maryland, College Park, MD 20742}}
\author{C.~Howell}
\affiliation{\mbox{Duke University and TUNL, Durham, NC 27708}}
\author{T.~Hunyady}
\affiliation{\mbox{Old Dominion University, Norfolk, VA 23529}}
\author{C.~E.~Hyde-Wright}

\affiliation{\mbox{Old Dominion University, Norfolk, VA 23529}}
\author{X.~Jiang}
\affiliation{\mbox{Rutgers, The State University of New Jersey\tc Piscataway, NJ 08854}}
\author{M.~K.~Jones}
\affiliation{\mbox{Thomas Jefferson National Accelerator Facility\tc Newport News, VA 23606}}
\author{M.~Khandaker}
\affiliation{\mbox{Norfolk State University, Norfolk, VA 23504}}
\author{A.~Ketikyan}
\affiliation{\mbox{Yerevan Physics Institute, Yerevan 375036, Armenia}}
\author{V.~Koubarovski}
\affiliation{\mbox{Rensselaer Physics Institute, Troy, NY 12180}}
\author{K.~Kramer}
\affiliation{\mbox{College of William and Mary, Williamsburg, VA 23187}}
\author{G.~Kumbartzki}
\affiliation{\mbox{Rutgers, The State University of New Jersey\tc Piscataway, NJ 08854}}
\author{G.~Laveissi\`ere}
\affiliation{\mbox{Universit\'{e} Blaise Pascal/IN2P3\tc F-63177 Aubi\`{e}re, France}}
\author{J.~LeRose}
\affiliation{\mbox{Thomas Jefferson National Accelerator Facility\tc Newport News, VA 23606}}
\author{R.~A.~Lindgren}
\affiliation{\mbox{University of Virginia, Charlottesville, VA 22901}}
\author{D.~J.~Margaziotis}
\affiliation{\mbox{California State University Los Angeles\tc Los Angeles\tc CA 90032}}
\author{P.~ Markowitz}
\affiliation{\mbox{Florida International University, Miami, FL
33199}}
\author{K.~McCormick}
\affiliation{\mbox{Old Dominion University, Norfolk, VA 23529}}
\author{Z.-E.~Meziani}
\affiliation{\mbox{Temple University, Philadelphia, PA 19122}}
\author{R.~Michaels}
\affiliation{\mbox{Thomas Jefferson National Accelerator Facility\tc Newport News, VA 23606}}
\author{P.~Moussiegt}
\affiliation{\mbox{Institut des Sciences Nucleiares\tc CNRS-IN2P3\tc F-38016 Grenoble, France}}
\author{S.~Nanda}
\affiliation{\mbox{Thomas Jefferson National Accelerator Facility\tc Newport News, VA 23606}}
\author{A.~M.~Nathan}
\affiliation{\mbox{University of Illinois, Urbana-Champaign, IL 61801}}
\author{D.~M.~Nikolenko}
\affiliation{\mbox{Budker Institute for Nuclear Physics\tc Novosibirsk 630090, Russia}}
\author{V.~Nelyubin}
\affiliation{\mbox{ St.~Petersburg Nuclear Physics Institute\tc Gatchina, 188350, Russia}}
\author{B.~E.~Norum}
\affiliation{\mbox{University of Virginia, Charlottesville, VA 22901}}
\author{K.~Paschke}
\affiliation{\mbox{University of Massachusetts, Amherst, MA 01003}}
\author{L.~Pentchev}
\affiliation{\mbox{College of William and Mary, Williamsburg, VA 23187}}
\author{C.~F.~Perdrisat}
\affiliation{\mbox{College of William and Mary, Williamsburg, VA 23187}}
\author{E.~Piasetzky}
\affiliation{\mbox{Tel Aviv University, Tel Aviv 69978, Israel}}
\author{R.~Pomatsalyuk}
\affiliation{\mbox{Kharkov Insitute of Physics and Technology\tc Kharkov 61108, Ukraine}}
\author{V.~A.~Punjabi}
\affiliation{\mbox{Norfolk State University, Norfolk, VA 23504}}
\author{I.~Rachek}
\affiliation{\mbox{Budker Institute for Nuclear Physics\tc Novosibirsk 630090, Russia}}
\author{A.~Radyushkin}
\affiliation{\mbox{Thomas Jefferson National Accelerator Facility\tc Newport News, VA 23606}}
\affiliation{\mbox{Old Dominion University, Norfolk, VA 23529}}
\author{B.~Reitz}
\affiliation{\mbox{Thomas Jefferson National Accelerator Facility\tc Newport News, VA 23606}}
\author{R.~Roche}
\affiliation{\mbox{Florida State University, Tallahassee, FL 32306}}
\author{M.~Roedelbronn}
\affiliation{\mbox{University of Illinois, Urbana-Champaign, IL 61801}}
\author{G.~Ron}
\affiliation{\mbox{Tel Aviv University, Tel Aviv 69978, Israel}}
\author{F.~Sabatie}
\affiliation{\mbox{Old Dominion University, Norfolk, VA 23529}}
\author{A.~Saha}
\affiliation{\mbox{Thomas Jefferson National Accelerator Facility\tc Newport News, VA 23606}}
\author{N.~Savvinov}
\affiliation{\mbox{University of Maryland, College Park, MD 20742}}
\author{A.~Shahinyan}
\affiliation{\mbox{Yerevan Physics Institute, Yerevan 375036, Armenia}}
\author{Y.~Shestakov}
 \affiliation{\mbox{Budker Institute for Nuclear Physics\tc Novosibirsk 
630090, Russia}}
\author{S.~\v{S}irca}
\affiliation{\mbox{Massachusetts Institute of Technology, Cambridge, MA 02139}}
\author{K.~Slifer}
\affiliation{\mbox{Temple University, Philadelphia, PA 19122}}
\author{P.~Solvignon}
\affiliation{\mbox{Temple University, Philadelphia, PA 19122}}
\author{P.~Stoler}
\affiliation{\mbox{Rensselaer Physics Institute, Troy, NY 12180}}
\author{S.~Tajima}
\affiliation{\mbox{Duke University and TUNL, Durham, NC 27708}}
\author{V.~Sulkosky}
\affiliation{\mbox{College of William and Mary, Williamsburg, VA 23187}}
\author{L.~Todor}
\affiliation{\mbox{Old Dominion University, Norfolk, VA 23529}}
\author{B.~Vlahovic}
\affiliation{\mbox{North Carolina Central University, Durham, NC 27707}}
\author{L.~B.~Weinstein}
\affiliation{\mbox{Old Dominion University, Norfolk, VA 23529}}
\author{K.~Wang}
\affiliation{\mbox{University of Virginia, Charlottesville, VA 22901}}
\author{B.~Wojtsekhowski}
\affiliation{\mbox{Thomas Jefferson National Accelerator Facility\tc Newport News, VA 23606}}
\author{H.~Voskanyan}
\affiliation{\mbox{Yerevan Physics Institute, Yerevan 375036, Armenia}}
\author{H.~Xiang}
\affiliation{\mbox{Massachusetts Institute of Technology, Cambridge, MA 02139}}
\author{X.~Zheng}
\affiliation{\mbox{Massachusetts Institute of Technology, Cambridge, MA 02139}}
\author{L.~Zhu}
\affiliation{\mbox{Massachusetts Institute of Technology, Cambridge, MA 02139}}

\collaboration{The Jefferson Lab Hall A Collaboration}
\noaffiliation{}

\date{\today}

\begin{abstract}                
Compton scattering from the proton was investigated at 
\invs~=~6.9~\gevsq~and \invt~=~-4.0~\gevsq~via polarization transfer 
from circularly polarized incident photons.
The longitudinal and transverse components of the recoil proton polarization
were measured.
The results are in excellent agreement with a prediction based on a reaction 
mechanism in which the photon interacts with a single
quark carrying the spin of the proton and in disagreement with a
prediction of pQCD based on a two-gluon exchange mechanism.
\end{abstract}

\preprint{APS/123-QED}

\maketitle

Real Compton Scattering (RCS) from the nucleon with $s$, $-t$, and $-u$ 
values large compared to $\Lambda_{_{QCD}}^2$ is a hard
exclusive process which provides access to information about
nucleon structure complementary to high Q$^2$ elastic form factors
\cite{an94,ga02} and Deeply Virtual Compton Scattering \cite{dvcs}.  A
common feature of these reactions is a high energy scale, leading
to factorization of the scattering amplitude into a hard
perturbative amplitude, which describes the coupling of the
external particles to the active quarks, and the overlap of soft
nonperturbative wave functions.

\begin{figure}[htb]
\epsfig{file=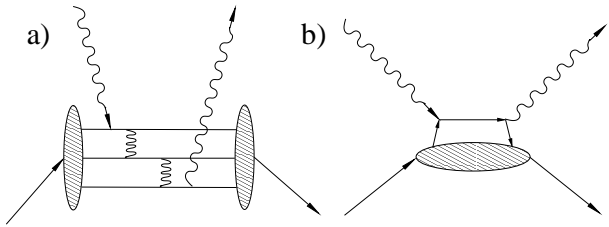,angle=0,width=8cm} \caption{RCS diagrams
for the (a) pQCD and (b) handbag reaction mechanism.}
\label{fig:diagram}
\end{figure}

Various theoretical approaches have been applied to RCS in the
hard scattering regime, and these can be distinguished by the
number of active quarks participating in the hard scattering
subprocess, or equivalently, by the mechanism for sharing the
transferred momentum among the constituents.
Two extreme pictures have been proposed.
In the perturbative QCD (pQCD) approach
(Fig.~\ref{fig:diagram}a) \cite{br81,fa90,kr91,br00},
three active quarks share the transferred momentum by the
exchange of two hard gluons. In the handbag approach
(Fig.~\ref{fig:diagram}b)\cite{ra98,di99,hu02,mi04}, there is only one
active quark whose wave function has sufficient high-momentum
components for the quark to absorb and re-emit the photon.  
In any given kinematic regime, both mechanisms will contribute, in
principle, to the cross section. It is generally believed that at
sufficiently high energies, the pQCD mechanism dominates. However,
the question of how high is ``sufficiently high'' is still open,
and it is not known with any certainty whether the pQCD mechanism
dominates in the kinematic regime that is presently accessible
experimentally.

One prediction of the pQCD mechanism for RCS is the constituent
scaling rule \cite{br73}, whereby $d\sigma/dt$ scales as $s^{-6}$
at fixed $\theta_{CM}$.
The only data in the few GeV regime from the pioneering experiment
at Cornell \cite{sh79} are approximately consistent
with constituent scaling, albeit with modest statistical precision.
Nevertheless, detailed calculations
show that the pQCD cross section underpredicts the data by
factors of at least ten \cite{kr91}, thereby calling into question
the applicability of the pQCD mechanism in this energy range.
On the other hand, more recent calculations using the handbag
approach have reproduced the Cornell cross-section data to better 
than a factor of two \cite{ra98,di99}.
The purpose of the present experiment~\cite{hy99} was to provide
a more stringent test of the reaction mechanism by improving
significantly on the statistical precision of the Cornell data,
by extending those data over a broader kinematic range, and by
measuring the polarization transfer observables \KLL~ and \KLS~at
a single kinematic point.
The results of the latter measurements are reported in this Letter.
As will be shown subsequently, these results are in unambiguous
agreement with the handbag mechanism and in disagreement
with the pQCD mechanism.

\begin{figure}[htb]
\epsfig{file=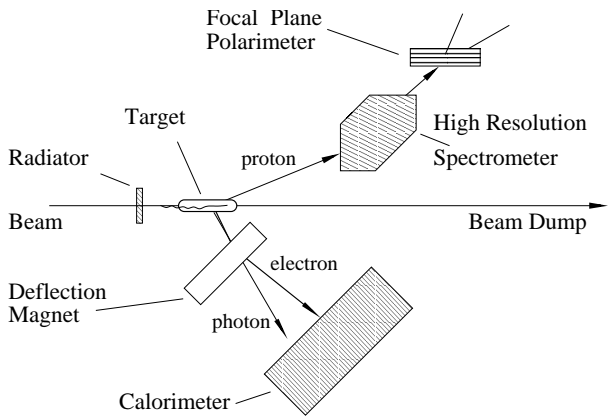,angle=0,width=8cm}
\caption{Schematic layout of the RCS experiment.}
\label{fig:scheme}
\end{figure}

The present measurement, shown schematically in
Fig.~\ref{fig:scheme}, was carried out in Hall A at Jefferson Lab,
with basic instrumentation described in \cite{exp:NIM}. A
longitudinally-polarized, 100\% duty-factor electron beam with
current up to 40 $\mu$A and energy of 3.48 GeV was incident on
a Cu radiator of 0.81 mm thickness.
The mixed beam of electrons and bremsstrahlung
photons was incident on a 15-cm liquid H$_2$ target, located just
downstream of the radiator, with a photon flux of up to $2 \times
10^{13}$ equivalent quanta/s.
Quasi-real photons, which contribute 16\% of total events with
an average virtuality of 0.005 GeV$^2$, were treated as 
a part of the RCS event sample.
For incident photons at a mean energy of 3.22 GeV, the scattered
photon was detected at a mean scattering angle of 65$^\circ$
in a calorimeter consisting of an array of 704 lead-glass blocks 
subtending a solid angle of 30 msr and with angular resolution 
of 1.8 mrad and relative energy resolution of 7.7\%.
The associated recoil proton was detected in one of the Hall A
High Resolution Spectrometers (HRS) at the corresponding central
angle of 20$^\circ$ and central momentum of 2.94 GeV.
The HRS had a solid angle of 6.5 msr, momentum
acceptance of $\pm$4.5\%, relative momentum resolution of $2.5 \times
10^{-4}$, and angular resolution of 2.4~mrad, the latter limited
principally by scattering in the target.
The trigger was formed from a coincidence between a signal from a
scintillator counter in the HRS focal plane and
a signal above a 500 MeV threshold in the calorimeter.
In total, 15 C and 3.5 C of beam charge were
accumulated for RCS production and calibration runs, respectively.

\begin{figure}[htb]
\epsfig{file=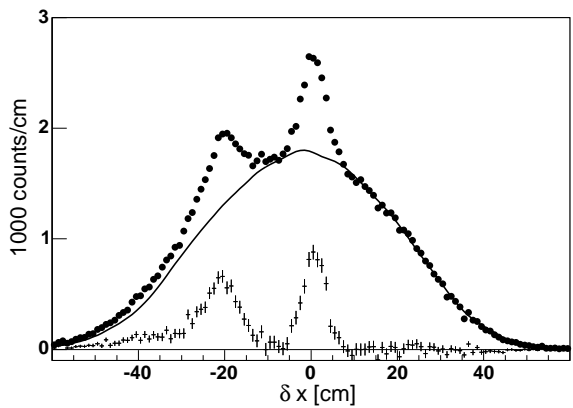,angle=0,width=8cm} 
\caption{Distribution of events in $\delta x$ (points) with a
coplanarity cut $\mid\delta y\mid \leqslant$ 10 cm.
The RCS and $ep$ elastic events form the peaks at $\delta x$ = 0
and -20 cm, respectively, and the photo-pion events the
underlying continuum.
A Monte Carlo simulation of the latter events is indicated by 
the solid curve.
The difference between the data and simulation is shown by 
the points with uncertainties.} 
\label{fig:dxdy}
\end{figure}

Potential RCS events were selected based on the kinematic
correlation between the scattered photon and the recoil proton.
The excellent HRS optics was used to reconstruct the momentum,
direction, and reaction vertex of the recoil proton and to
calculate $\delta x$ and $\delta y$, the difference in $x$ and $y$
coordinates, respectively, between the expected and measured
location of the detected photon on the front face of the
calorimeter.
The distribution of events in $\delta x$ with a
coplanarity cut of $\mid\delta y\mid \leqslant$ 10 cm is shown in
Fig.~\ref{fig:dxdy}.
The RCS events, which are in the peak at $\delta x=0$, lie upon
a continuum background primarily from the $p(\gamma,\pi^0 p)$ reaction,
with the subsequent decay $\pi^0\rightarrow\gamma\gamma$.
An additional background is due to electrons from $ep$ elastic
scattering, which is kinematically indistinguishable from RCS.
A magnet between the target and the calorimeter
(see Fig.~\ref{fig:scheme}) deflected these electrons horizontally
by $\sim$20 cm relative to undeflected RCS photons.
%

The recoil proton polarization was measured by the focal plane
polarimeter (FPP) located in the HRS.  The FPP determines the two
polarization components normal to the momentum of the proton by
measuring the azimuthal asymmetries in the angular distribution
after scattering the proton from an analyzer, then taking the difference 
of these asymmetries between plus and minus electron beam helicity
states. To improve the efficiency, two analyzers were utilized in
the experiment, a 44-cm block of CH$_2$ and a 50-cm block of carbon. 
Vertical drift chambers together with front and rear
straw chambers tracked the protons before, between, and after the
analyzers, effectively producing two independent polarimeters with
a combined product of efficiency and square of analyzing power
that was measured to be $4.5\times10^{-3}$.
For each analyzer separately, Fourier analysis of the helicity
difference leads to the product of the proton polarizations at the
FPP ($P_n^{fpp}$ or $P_t^{fpp}$), the circular polarization of the
incident photon beam ($P^c_\gamma$), and the FPP analyzing power
($A_y$):
\begin{center}
$N \left( \vartheta, \varphi \right) = N_{0} \left( \vartheta
\right) \left\{ 1 + \left[ P^c_\gamma A_y \left( \vartheta \right)
P_t^{fpp} + \alpha \right] \sin \varphi \right. $ \\
$ \left. - \left[ P^c_\gamma A_y \left( \vartheta \right) P_n^{fpp}
+ \beta \right] \cos \varphi \right\} $, \\
\end{center}
where  $N_{0}$ is the number of protons which scatter in the polarimeter,
$\vartheta$ and $\varphi$ are the polar and azimuthal scattering angles, 
and $\alpha,\beta$ are instrumental asymmetries.
Determination of $A_{y}(\vartheta)$, $\alpha$, and $\beta$ 
for each analyzer was performed by measuring the polarization of 
the recoil proton from $\vec{e}p$ elastic scattering at approximately 
the same momentum and by using previously determined ratio of 
the proton form factors~\cite{ga02}.
The electron beam polarization was measured to be $0.766\pm0.026$
at the start of the experiment using a M\o{}ller polarimeter and
continuously monitored throughout the production runs by observing
the asymmetry due to the large $p(\gamma,\pi^0 p)$ background.  
An upper limit of 2\% for the change of the beam polarization 
during the experiment was obtained from the pion data.
The bremsstrahlung photon has 99\% of the initial electron
polarization over the energy range used~\cite{ols59}.

To relate the proton polarization components at the focal plane to
their counterparts at the target, the precession of the proton
spin in the HRS magnetic elements was taken into account using a
COSY model~\cite{cosy99} of the HRS optics for the spin transport
matrix.
The elements of this matrix depend on the total precession angle, which 
was near $270^\circ$ in order to optimize the determination of \KLL.
The proton spin vector was then transformed to the proton rest frame,
with the longitudinal axis pointing in the direction of the recoil
proton in the center of mass frame~\cite{di99}.
In that frame, the longitudinal and transverse components of 
the proton polarization are just the spin transfer parameters 
\KLL~and \KLS, respectively.

The RCS events are selected from a small elliptical region at the
origin of the $\delta x-\delta y$ plane. For each spin component,
the RCS recoil polarization is given by
$$
P_{_{RCS}}\,=\,\left [P_{all}-(1-R)P_{bkg}\right ]/R \, ,
 \label{eq:prcs}
$$
where $P_{all}$ and $P_{bkg}$ are the polarizations for all events
and background events in that region, respectively, and R is the
ratio of RCS to total events.
The background polarization was measured by selecting events from
regions of the $\delta x-\delta y$ plane that contain neither RCS
nor $ep$ elastic events.
It was determined that within the statistical precision of the
measurements, $P_{bkg}$ was constant over broad regions of that plane.
Results obtained with the two polarimeters were
statistically consistent and were averaged.
With the RCS region selected to obtain the best accuracy on
$P_{_{RCS}}$, one finds $R=0.383\pm 0.004$ 
and the resulting polarizations are given in Table~\ref{tab:results}.

\begin{table}[htb]
\caption{Proton recoil polarizations.  For the RCS polarization
the first uncertainty is statistical; the second is systematic and
dominated by the background subtraction.}
\begin{ruledtabular}
\renewcommand{\arraystretch}{1.5}
\begin{tabular}{cccc}
  & $P_{all}$& $P_{bkg}$ & $P_{_{RCS}}$=\KLL~(\KLS) \\
\hline
LL & 0.588$\pm$0.030     & 0.532$\pm$0.006& 0.678$\pm$0.083$\pm$0.04 \\
LS & 0.340$\pm$0.029     & 0.480$\pm$0.006& 0.114$\pm$0.078$\pm$0.04 
\end{tabular}
\end{ruledtabular}
\label{tab:results}
\end{table}

\begin{figure}[tb]
\epsfig{file=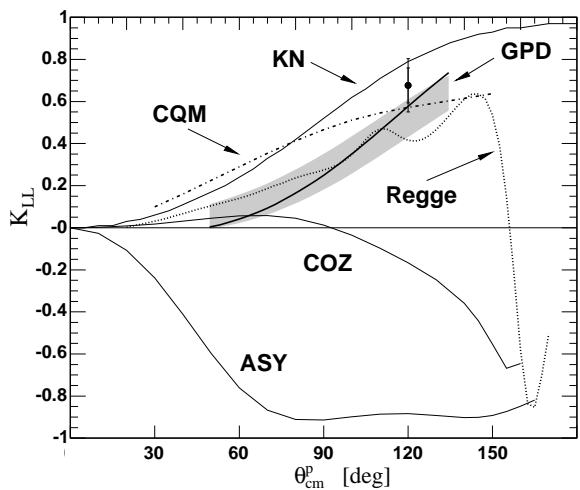,angle=0,width=8cm} \caption{Our result
for \KLL~compared with calculations in different approaches:
ASY and COZ both from pQCD \cite{br00}, GPD \cite{hu02},
CQM \cite{mi04}, and extended Regge model \cite{ca03}.
The curve labeled KN is $K_{_{LL}}^{^{KN}}$, 
the Klein-Nishina asymmetry for a structureless proton.}
\label{fig:kll}
\end{figure}

The result for \KLL~ is shown in Fig.~\ref{fig:kll} along with
the results of relevant calculations.
In the handbag calculation using Generalized Parton Distributions (GPD),
\begin{center}
$K_{_{LL}} \simeq \frac{R_{_A}}{R_{_V}} \, K_{_{LL}}^{^{KN}}$
$\left[1 - \frac{t^2}{2(s^2 + u^2)} \left(1 - 
\frac{R^2_{_A}}{R^2_{_V}} \right) \right]^{-1}$ 
\end{center}
where $R_{_A},R_{_V}$ are axial and vector form factors,
respectively, that are unique to the RCS process \cite{di99}.
The experimental result implies the ratio 
${R_{_A}}/{R_{_V}}$ = $0.81 \pm 0.11$.

The excellent agreement between the experiment and the GPD-based 
calculation, shown with a range of uncertainties due to finite 
mass corrections~\cite{di03}, and the close proximity of each to
$K_{_{LL}}^{^{KN}}$ are consistent with a picture in which the 
photon scatters from a single quark whose spin is in the direction 
of the proton spin.
The RCS form factors are certain moments of the GPD's $H$
and $\tilde{H}$ \cite{ra98,di99}, 
so our result provides a constraint on relative values of 
these moments.
An alternative handbag-type approach using constituent quarks 
(CQM)~\cite{mi04}, with parameters adjusted to fit \gep~data 
\cite{ga02}, is also in excellent agreement with the datum.
Also in good agreement is a semi-phenomenological calculation using
the extended Regge model~\cite{ca03}, with parameters fixed by
a fit to high-$t$ photoproduction of vector mesons.
On the other hand, the pQCD calculations~\cite{br00}, shown for
both the asymptotic (ASY) and the COZ \cite{ch89} distribution 
amplitude, disagree strongly with the experimental point, 
suggesting that the asymptotic regime has not yet been reached.

A non-zero value of $K_{_{LS}}$ implies a proton helicity-flip process,
which is strictly forbidden in leading-twist pQCD.
In the GPD-based approach~\cite{hu02}, 
$K_{_{LS}}/K_{_{LL}} \simeq ({\sqrt{-\hat{t}}}/{2M}) R_{_T}/R_{_V}$, 
where $\hat{t}$ is the four-momentum transfer in the hard subprocess
of the handbag diagram, $M$ is the proton mass, 
and $R_{_T}$ is a tensor form factor of the RCS process.
From the experimental result for $K_{_{LS}}$,
we estimate $R_{_T}/R_{_V} = 0.21 \pm 0.11 \pm 0.03$,
where the first uncertainty is statistical and the second is
systematic due to the mass correction uncertainty 
in calculating $\hat{t}$ \cite{di03}. 
A value of 0.33 was predicted for $R_{_T}/R_{_V}$ \cite{hu02}
based on the hypothesis $R_{_T}/R_{_V} = F_2/F_1$, the ratio of 
the Dirac and the Pauli electromagnetic form factors.
Although the uncertainties are large, the present data suggest
that $R_{_T}/R_{_V}$ may fall more rapidly with $-t$ than
$F_2/F_1$.  
$K_{_{LS}}$ vanishes in the CQM-based handbag
calculation \cite{mi04}.
                  
In conclusion, the polarization transfer observables \KLL~ and
\KLS~were measured for proton Compton scattering in the wide-angle
regime at \mbox{\invs~=6.9}, \mbox{\invt~=-4.0~\gevsq}
and shown to be in good agreement with calculations based on the
handbag reaction mechanism \cite{hu02,mi04}.

We thank the Jefferson Lab Hall A technical staff for their outstanding
support.  This work was supported in part from the National Science
Foundation, the UK Engineering and Physical Science Research Council, 
and the DOE under contract DE-AC05-84ER40150 Modification 
No. M175, under which the Southeastern Universities Research Association 
(SURA) operates the Thomas Jefferson National Accelerator Facility.

\end{document}